\documentclass[letter]{aa}

\usepackage{graphicx}
\usepackage{txfonts}
\usepackage{lipsum}
\usepackage{subcaption}         % necessary for continued figures, example in section 3
\usepackage{lscape}             % to rotate a single page table, example in appendix.
\usepackage{placeins}           % useful with \FloatBarrier, to keep 
\usepackage{array}              % Table formatting
\usepackage{multirow}
\usepackage{soul}
\usepackage{xcolor}
\usepackage{orcidlink}
\newcommand{\orcid}[1]{}              % hide ORCIDs for A&A submission

\makeatletter
\newif\ifaa@deferaffil \aa@deferaffiltrue
\let\aa@origaainstitutename\aainstitutename
\renewcommand*\aainstitutename{%
  \ifaa@deferaffil
    {\noindent\aa@longauthfont\aa@longauthopen\aa@longauthname\aa@longauthclose\par}%
  \else
    \aa@origaainstitutename
  \fi}
\newcommand*\printaffiliations{%
  \aa@deferaffilfalse
  \par\nolinenumbers
  \kern6pt\hrule\kern6pt 
  \aainstitutename\par}
\makeatother

\defcitealias{DES:2026zjp}{Sanchez-Cid, Fert\'e et al. 2026}

\usepackage{hyperref}
\hypersetup{
    colorlinks=true,
    linkcolor=blue,
    urlcolor=blue,
    citecolor=blue,
    }

\usepackage{acronym}
\newacro{CDM}[CDM]{cold dark matter}
\newacro{EM}[EM]{electromagnetic}
\newacro{WL}[WL]{weak lensing}
\newacro{GC}[GC]{galaxy clustering}
\newacro{DES}[DES]{Dark Energy Survey}
\newacro{BAO}[BAO]{baryon acoustic oscillations}
\newacro{SNe Ia}[SNe Ia]{type Ia supernovae}
\newacro{SPT}[SPT]{South Pole Telescope}
\newacro{ACT}[ACT]{Atacama Cosmology Telescope}
\newacro{DR6}[DR6]{data release 6}
\newacro{DESI}[DESI]{Dark Energy Spectroscopic Instrument}
\newacro{DR2}[DR2]{data release 2}
\newacro{GWTC-4.0}[GWTC-4.0]{fourth Gravitational-Wave Transient Catalog} 
\newacro{GWTC-5.0}[GWTC-5.0]{fifth Gravitational-Wave Transient Catalog} 
\newacro{GW}[GW]{gravitational wave}
\newacro{LVK}[LVK]{LIGO-Virgo-KAGRA Collaboration}
\newacro{CBC}[CBC]{compact binary coalescence}
\newacro{PE}[PE]{parameter estimation}
\newacro{FAR}[FAR]{false alarm rate}
\newacro{BBH}[BBH]{binary black hole}
\newacro{CMB}[CMB]{cosmic microwave background}
\newacro{MAP}[MAP]{maximum a posteriori}
\newacro{APS}[APS]{angular power spectra}
\newacro{2PCF}[2PCF]{two-point correlation function}
\newacro{KiDS}[KiDS]{Kilo-Degree Survey}
\newacro{HSC}[HSC]{Hyper-Suprime Camera}
\newacro{LSST}[LSST]{Legacy Survey of Space and Time}
\acrodef{LIGO}[LIGO]{Laser Interferometer Gravitational-Wave Observatory}
\newacro{NF}[NF]{normalizing flows}
\newacro{IS}[IS]{importance sampling}
\newacro{FoM}[FoM]{figure-of-merit}
\newacro{MCMC}[MCMC]{Markov Chain Monte Carlo}
\newacro{BBN}[BBN]{Big Bang nucleosynthesis}

\newif\ifannotations
\annotationsfalse

\ifannotations

  \newcommand{\lvkir}[1]{\textcolor{red}{LVK-IR: #1}}
  \newcommand{\desirjm}[1]{\textcolor{orange}{DES-JM: #1}}
  \newcommand{\ver}[1]{\textcolor{pink}{v2: #1}}
\else

  \newcommand{\lvkir}[1]{#1}
  \newcommand{\desirjm}[1]{#1}
  \newcommand{\ver}[1]{#1}
\fi

\makeatletter
\renewcommand{\@fnsymbol}[1]{\ensuremath{\ifcase#1\or\dagger\or\ddagger\or\star\or\star\star\fi}}
\makeatother

\usepackage{eso-pic}
\newcommand{\preprintnumbers}[1]{%
  \AddToShipoutPictureFG*{%
    \AtPageUpperLeft{%
      \put(\LenToUnit{\dimexpr1in+\oddsidemargin+\textwidth\relax},\LenToUnit{-1.15in}){%
        \makebox[0pt][r]{\small\begin{tabular}{@{}r@{}}#1\end{tabular}}}}}}

\begin{document}

\preprintnumbers{DES-2026-0983\\FERMILAB-PUB-26-0788}

%%%%%%%%%%%%%%%%%%%%%%%%%%%%%%%%%%%%%%%%
% if you use custom commands in your title,
% ensure to check your title when submitting!
%%%%%%%%%%%%%%%%%%%%%%%%%%%%%%%%%%%%%%%%
   \title{A multiprobe $\Lambda$CDM consistency analysis with gravitational-wave standard sirens}

%%%%%%%%%%%%%%%%%%%%%%%%%%%%%%%%%%%%%%%%
% Please separate each author with the \and command
%
% Please do not include ORCIDs next to author names.
% Only ORCIDs authenticated by individual authors in EDPS
% editorial system will be taken into account.
% ORCIDs included here will be removed.
%%%%%%%%%%%%%%%%%%%%%%%%%%%%%%%%%%%%%%%%

   \author{T.~M.~C.~Abbott\orcid{0000-0003-1587-3931}\inst{1}
        \and M.~Aguena\orcid{0000-0001-5679-6747}\inst{2,3}
        \and F.~Andrade-Oliveira\orcid{0000-0003-0171-6900}\inst{4}
        \and S.~Avila\orcid{0000-0001-5043-3662}\inst{5}
        \and D.~Bacon\inst{6}
        \and D.~Brooks\orcid{0000-0002-8458-5047}\inst{7}
        \and H.~Camacho\inst{3,8}
        \and R.~Camilleri\inst{9}
        \and G.~Campailla\inst{10}
        \and A.~Carnero~Rosell\orcid{0000-0003-3044-5150}\inst{3,11,12}
        \and J.~Carretero\orcid{0000-0002-3130-0204}\inst{13}
        \and R.~Cawthon\orcid{0000-0003-2965-6786}\inst{14}
        \and C.~Chang\orcid{0000-0002-7887-0896}\inst{15,16}
        \and L.~N.~da~Costa\orcid{0000-0002-7731-277X}\inst{3}
        \and M.~E.~da~Silva~Pereira\inst{17}
        \and S.~Desai\orcid{0000-0002-0466-3288}\inst{18}
        \and H.~T.~Diehl\orcid{0000-0002-8357-7467}\inst{19}
        \and P.~Doel\orcid{0000-0002-6397-4457}\inst{7}
        \and A.~Fert\'e\orcid{0000-0003-3065-9941}\inst{20}
        \and D.~L.~Hollowood\orcid{0000-0002-9369-4157}\inst{21}
        \and K.~Honscheid\orcid{0000-0002-6550-2023}\inst{22,23}
        \and D.~Huterer\orcid{0000-0001-6558-0112}\inst{24}
        \and D.~Laghi\inst{4}
        \and S.~Lee\orcid{0000-0002-8289-740X}\inst{25,26}
        \and S.~MacBride\inst{4}
        \and J.~L.~Marshall\orcid{0000-0003-0710-9474}\inst{27}
        \and I.~McMahon\orcid{0000-0002-4529-1505}\inst{4}
        \and J.~Mena-Fern{\'a}ndez\orcid{0000-0001-9497-7266}\inst{28,29}
        \and R.~Miquel\orcid{0000-0002-6610-4836}\inst{13,30}
        \and J.~Muir\orcid{0000-0002-7579-770X}\inst{31}
        \and J.~Myles\orcid{0000-0001-6145-5859}\inst{32}
        \and R.~L.~C.~Ogando\orcid{0000-0003-2120-1154}\inst{33,34}
        \and A.~A.~Plazas~Malag\'on\orcid{0000-0002-2598-0514}\inst{20,35}
        \and J.~Prat\orcid{0000-0002-5933-5150}\inst{36,37}
        \and M.~Rodr\'{i}guez-Monroy\inst{38}
        \and R.~Rosenfeld\inst{3,39}
        \and S.~Samuroff\orcid{0000-0001-7147-8843}\inst{13,40}
        \and E.~Sanchez\orcid{0000-0002-9646-8198}\inst{5}
        \and D.~Sanchez-Cid\orcid{0000-0003-3054-7907}\inst{4}\fnmsep\thanks{\email{david.sanchezcid@physik.uzh.ch}}
        \and I.~Sevilla-Noarbe\orcid{0000-0002-1831-1953}\inst{5}
        \and M.~Soares-Santos\orcid{0000-0001-6082-8529}\inst{4}
        \and E.~Suchyta\orcid{0000-0002-7047-9358}\inst{41}
        \and M.~E.~C.~Swanson\inst{42}
        \and V.~Vikram % TODO: no affiliation in DES-2026-0983 author list
        \and N.~Weaverdyck\orcid{0000-0001-9382-5199}\inst{43,44}
        \and J.~Weller\orcid{0000-0002-8282-2010}\inst{45,46}
        \and M.~Yamamoto\orcid{0000-0003-1585-997X}\inst{32,47}
        }

   \institute{Cerro Tololo Inter-American Observatory, NSF's National Optical-Infrared Astronomy Research Laboratory, Casilla 603, La Serena, Chile
        \and INAF-Osservatorio Astronomico di Trieste, via G. B. Tiepolo 11, I-34143 Trieste, Italy
        \and Laborat\'orio Interinstitucional de e-Astronomia - LIneA, Av. Pastor Martin Luther King Jr, 126 Del Castilho, Nova Am\'erica Offices, Torre 3000/sala 817 CEP: 20765-000, Brazil
        \and Physik-Institut, University of Zürich, Winterthurerstrasse 190, CH-8057 Zürich, Switzerland
        \and Centro de Investigaciones Energ\'eticas, Medioambientales y Tecnol\'ogicas (CIEMAT), Madrid, Spain
        \and Institute of Cosmology and Gravitation, University of Portsmouth, Portsmouth, PO1 3FX, UK
        \and Department of Physics \& Astronomy, University College London, Gower Street, London, WC1E 6BT, UK
        \and Brookhaven National Laboratory, Bldg 510, Upton, NY 11973, USA
        \and School of Mathematics and Physics, University of Queensland, Brisbane, QLD 4072, Australia
        \and Department of Physics, University of Genova and INFN, Via Dodecaneso 33, 16146, Genova, Italy
        \and Instituto de Astrofisica de Canarias, E-38205 La Laguna, Tenerife, Spain
        \and Universidad de La Laguna, Dpto. Astrofísica, E-38206 La Laguna, Tenerife, Spain
        \and Institut de F\'{\i}sica d'Altes Energies (IFAE), The Barcelona Institute of Science and Technology, Campus UAB, 08193 Bellaterra (Barcelona) Spain
        \and Oxford College of Emory University, Oxford, GA 30054, USA
        \and Department of Astronomy and Astrophysics, University of Chicago, Chicago, IL 60637, USA
        \and Kavli Institute for Cosmological Physics, University of Chicago, Chicago, IL 60637, USA
        \and Department of Physics and Astronomy, Pevensey Building, University of Sussex, Brighton, BN1 9QH, UK
        \and Department of Physics, IIT Hyderabad, Kandi, Telangana 502285, India
        \and Fermi National Accelerator Laboratory, P. O. Box 500, Batavia, IL 60510, USA
        \and SLAC National Accelerator Laboratory, Menlo Park, CA 94025, USA
        \and Santa Cruz Institute for Particle Physics, Santa Cruz, CA 95064, USA
        \and Center for Cosmology and Astro-Particle Physics, The Ohio State University, Columbus, OH 43210, USA
        \and Department of Physics, The Ohio State University, Columbus, OH 43210, USA
        \and Department of Physics, University of Michigan, Ann Arbor, MI 48109, USA
        \and Department of Physics and Astronomy, Ohio University, Clippinger Labs, Athens, OH 45701, USA
        \and Jet Propulsion Laboratory, California Institute of Technology, 4800 Oak Grove Dr., Pasadena, CA 91109, USA
        \and George P. and Cynthia Woods Mitchell Institute for Fundamental Physics and Astronomy, and Department of Physics and Astronomy, Texas A\&M University, College Station, TX 77843, USA
        \and Aix Marseille Univ, CNRS/IN2P3, CPPM, Marseille, France
        \and Universit\'e Grenoble Alpes, CNRS, LPSC-IN2P3, 38000 Grenoble, France
        \and Instituci\'o Catalana de Recerca i Estudis Avan\c{c}ats, E-08010 Barcelona, Spain
        \and Department of Physics, University of Cincinnati, Cincinnati, Ohio 45221, USA
        \and Department of Astrophysical Sciences, Princeton University, Peyton Hall, Princeton, NJ 08544, USA
        \and Centro de Tecnologia da Informa\c{c}\~ao Renato Archer, Campinas, SP, Brazil - 13069-901
        \and Observat\'orio Nacional, Rio de Janeiro, RJ, Brazil - 20921-400
        \and Kavli Institute for Particle Astrophysics \& Cosmology, P. O. Box 2450, Stanford University, Stanford, CA 94305, USA
        \and Departament de F\'isica, EEBE, Universitat Polit\`ecnica de Catalunya, c/ Eduard Maristany 16, E-08019 Barcelona, Spain
        \and University of Copenhagen, Dark Cosmology Centre, Juliane Maries Vej 30, 2100 Copenhagen O, Denmark
        \and HUN-REN Csillagászati és Földtudományi Kutatóközpont, Konkoly-Thege Miklós út 15-17, Budapest 1121, Hungary
        \and ICTP South American Institute for Fundamental Research, Instituto de F\'{\i}sica Te\'orica, Universidade Estadual Paulista, S\~ao Paulo, Brazil
        \and Department of Physics, Northeastern University, Boston, MA 02115, USA
        \and Computer Science and Mathematics Division, Oak Ridge National Laboratory, Oak Ridge, TN 37831, USA
        \and Center for Astrophysical Surveys, National Center for Supercomputing Applications, 1205 West Clark St., Urbana, IL 61801, USA
        \and Berkeley Center for Cosmological Physics, Department of Physics, University of California, Berkeley, CA 94720, USA
        \and Lawrence Berkeley National Laboratory, 1 Cyclotron Road, Berkeley, CA 94720, USA
        \and Max Planck Institute for Extraterrestrial Physics, Giessenbachstrasse, 85748 Garching, Germany
        \and Universit\"ats-Sternwarte, Fakult\"at f\"ur Physik, Ludwig-Maximilians Universit\"at M\"unchen, Scheinerstr. 1, 81679 M\"unchen, Germany
        \and Department of Physics, Duke University, Durham, NC 27708, USA
        }

   \authorrunning{DES Collaboration} 
   \date{}

% \abstract{}{}{}{}{}
% 5 {} token are mandatory
 
\abstract{

Multi-messenger cosmology offers an independent path to stress-test the $\Lambda$CDM model as \ac{GW} catalogs grow in size. We present the first multiprobe assessment of $\Lambda$CDM tensions incorporating \ac{GW} standard sirens: \ac{LVK} \ac{GWTC-5.0} standard sirens are combined with \ac{DES} Year~6 weak lensing and galaxy clustering (3$\times$2pt), \ac{BAO}, and \ac{SNe Ia}, together with \ac{DESI} \ac{BAO}, and confronted with primary \ac{CMB} measurements from \textit{Planck}, the \ac{ACT}, and the \ac{SPT}. All probes are analyzed under homogeneous priors and combined via normalizing flows, and we quantify the $S_8$ discrepancy and $H_0$ tension with respect to \ac{CMB}. Adding standard sirens to the 3$\times$2pt analysis improves the constraints in the $\Omega_m$--$S_8$ plane by \ver{$\sim$30\%}, while constraining $H_0$ to \ver{6.2\%}. We present cosmological constraints from the low-redshift probes including standard sirens, yielding \ver{$S_8 = 0.793^{+0.010}_{-0.010}$, $\Omega_m = 0.308^{+0.007}_{-0.006}$, and $H_0 = 69.4^{+4.1}_{-4.4} \,\mathrm{km}\,\mathrm{s}^{-1}\,\mathrm{Mpc}^{-1}$}. At current \ac{GW} sensitivity, however, both the $S_8$ discrepancy and the $H_0$ tension remain essentially unchanged. Our analysis nonetheless delivers the first assessment of $\Lambda$CDM tensions computed with standard sirens included among the probes, and establishes them as a mature cosmological probe --- ready to be combined on equal footing with established datasets in forthcoming surveys such as the \ac{LSST}, \textit{Euclid}, and \textit{Roman}.
}


\maketitle
\nolinenumbers

\acresetall

\section{Introduction}
\label{sec:intro}

The $\Lambda$CDM model, comprising a cosmological constant ($\Lambda$) and \ac{CDM} that seeds large-scale structure, remains the standard framework for cosmic history and dynamics. Its robustness is supported by successful predictions tested against \ac{EM} observations: \ac{CMB}~\citep{Planck:2018vyg}, \ac{SNe Ia}~\citep{SupernovaSearchTeam:1998fmf, SupernovaCosmologyProject:1998vns}, and \ac{WL}~\citep{Refregier:2003ct}. However, increasing precision has revealed tensions within this framework: the $H_0$ tension at $5$--$7\sigma$~\citep{DiValentino:2021izs} between early-Universe probes and the local distance ladder from SH0ES~\citep{Riess:2021jrx}, and the mild but persistent discrepancy at $\lesssim 3\sigma$~\citep{Abdalla:2022yfr} between \ac{WL} and \ac{CMB} in the clustering amplitude $S_8 \equiv \sigma_8(\Omega_m/0.3)^{1/2}$. Since the first \ac{GW} detection \citep{LIGOScientific:2016aoc}, GW standard sirens have been discussed as a promising independent cosmological probe \citep{Schutz:1986gp, Holz:2005df, LIGOScientific:2021aug, McMahon:2026nhi, MacBride:2026ufy}, and first joint analyses with spectroscopic \ac{BAO} have recently been explored \citep{Pierra:2025hoc}, yet their combination with leading photometric datasets has yet to be carried out rigorously.

In this work, we fill this gap by combining \ac{LVK}~\ac{GWTC-5.0} standard sirens~\citep{LIGOScientific:2026uyd} with state-of-the-art \ac{EM} datasets and computing tension metrics in $S_8$ and $H_0$. For the first time, standard sirens are incorporated into the traditional low-$z$ combination of \ac{SNe Ia}, \ac{BAO}, and \ac{WL}, enabling a direct assessment of their contribution to current $\Lambda$CDM tensions.

\section{Data}
\label{sec:data}

\textbf{\acs{DES}} -- The \ac{DES} is a photometric galaxy survey covering one-eighth of the sky in \textit{grizY} bands using the Dark Energy Camera~\citep[DECam;][]{DES:2015wtr}. The legacy data release encompasses the full six-year dataset~\citep{DES:2025key}. The flagship measurement is the joint \ac{WL} and \ac{GC} analysis ($3\times2\mathrm{pt}$) of 140 million source galaxies~\citep{Yamamoto:2025jsd} up to $z \sim 2$ in four tomographic redshift bins, combined with 9 million lens galaxies~\citep{DES:2026aqv}; galaxy clustering is analyzed down to $4\,\mathrm{Mpc}/h$ using a non-linear galaxy bias model~\citetext{\citetalias{DES:2026zjp}; \citealt{desy6-3x2pt}}. Geometric constraints come from the \ac{BAO} measurement of 16 million luminous red galaxies at $0.6 < z < 1.2$ ($z_{\mathrm{eff}} = 0.85$)~\citep{DES:2024pwq, DES:2024wym}, with the \ac{DESI} overlap region excluded \desirjm{in a recent work~\citep{DES:2026aht}}. The dataset is completed with Year~5 \ac{SNe Ia}: \desirjm{1623} DES supernovae combined with \desirjm{197} external low-$z$ events~\citep{DES:2025sig}.

\textbf{\acs{DESI} \acs{BAO}} -- We use \ac{DESI} \ac{DR2} \ac{BAO} measurements~\citep{DESI:2025zgx} spanning $0.295 < z < 2.330$, comprising isotropic scaling from the BGS sample and both isotropic and Alcock--Paczy\'nski parameters from the LRG1, LRG2, LRG3+ELG1, ELG2, QSO, and Ly$\alpha$ tracers.

\textbf{\acs{LVK} \acs{GWTC-5.0}} -- The \ac{LVK} network~\citep{KAGRA:2013rdx} comprises four interferometers --- two \ac{LIGO}~\citep{LIGOScientific:2014pky} detectors in the USA, Virgo~\citep{VIRGO:2014yos} in Europe, and KAGRA~\citep{KAGRA:2020tym} in Japan --- sensitive to \acp{GW} in the ${\sim}10$--$1000$\,Hz band. The \ac{GWTC-5.0}~\citep{LIGOScientific:2026wfs} compiles \ac{GW} transients detected between the first observing run (O1) and the end of O4b (January 2025). As in~\citet{Andrade-Oliveira:2026jjm}, we use posterior samples for $H_0$ and $\Omega_m$ from the spectral siren analysis of 235 \acp{CBC} \lvkir{derived with \textsc{icarogw}~\citep{Mastrogiovanni:2023zbw}}; spectral sirens are \ac{GW} events whose source redshifts are inferred from the redshifted intrinsic population mass spectrum~\citep{Ezquiaga:2022zkx}, without requiring an \ac{EM} counterpart. GW170817 is the only bright siren in \ac{GWTC-5.0}; bright sirens are \ac{GW} events with an identified \ac{EM} counterpart that provides the source redshift directly. We supplement its \ac{GW} posterior with the host recession velocity $v = 3327 \pm 72$\,km\,s$^{-1}$~\citep{Crook:2006sw} relative to the \ac{CMB} and a radial peculiar-velocity correction $\langle v_p \rangle = 310 \pm 150$\,km\,s$^{-1}$~\citep{Carrick:2015xza}. We adopt the inclination-angle constraint $15^\circ < \theta_{\mathrm{JN}} < 29^\circ$, derived from superluminal jet motion observed 75--230 days post-merger~\citep{Mooley:2018qfh, Hotokezaka:2018dfi, Mukherjee:2019qmm}. More recent constraints on the jet properties~\citep{Gourdji:2026rqc} would tighten the $H_0$ posterior; we retain the more conservative modeling of~\citet{Andrade-Oliveira:2026jjm} as our baseline.

\textbf{\acs{CMB}} -- We combine \ac{CMB} temperature (T) and E-mode polarization (E) likelihoods from \textit{Planck}~\citep{Planck:2018vyg}, \ac{ACT} \ac{DR6}~\citep{AtacamaCosmologyTelescope:2025blo}, and the \acl{SPT} (SPT-3G)~\citep{SPT-3G:2025bzu} to obtain tight constraints on the full $\Lambda$CDM parameter space. From \textit{Planck} we use the \textsc{Plik-lite} likelihood ($2 < \ell_{\mathrm{TT}} < 1000$; $30 < \ell_{\mathrm{TE,EE}} < 600$)~\citep{Prince:2019hse} and \textsc{SimAll} ($2 < \ell_{\mathrm{EE}} < 30$) to constrain the optical depth to reionization $\tau$, where $\ell_{\mathrm{XY}}$ denotes the multipole range used for the XY spectrum. From \ac{ACT} \ac{DR6} we use the \textsc{lite} likelihood ($600 < \ell < 6500$; TT, TE, EE). From \acs{SPT}-3G we use the \textsc{lite} likelihood via \textsc{candl}~\citep{Balkenhol:2024sbv} ($400 < \ell_{\mathrm{TT}} < 3000$; $400 < \ell_{\mathrm{TE,EE}} < 4000$). We adopt the \textsc{lite} likelihood approximations which marginalize over foreground and calibration nuisances analytically.

\section{Methods}
\label{sec:methods}

\textbf{\acs{WL} + \acs{GC}} -- The $3\times2\mathrm{pt}$ analysis extracts cosmological information from \acp{2PCF} of the galaxy overdensity and shear fields: cosmic shear (shape--shape), galaxy clustering (position--position), and galaxy--galaxy lensing (position--shape). The theoretical framework accounts for small-scale systematics including baryonic feedback, non-linear galaxy bias, and intrinsic alignments; scales that are not accurately modeled are excluded from the fit~\citepalias{DES:2026zjp}. The Gaussian likelihood is
\begin{equation}
    \mathcal{L}(\mathrm{D} \mid \Theta) \propto 
    \exp\!\left\{ -\frac{1}{2} \mathrm{R}^\top \mathrm{Cov}^{-1} \mathrm{R} \right\},
\end{equation}
where $\mathrm{R} = \mathrm{D} - \mathrm{T}(\Theta)$, $\mathrm{D}$ is the data vector of 2PCFs, $\mathrm{T}(\Theta)$ the theory prediction, and $\mathrm{Cov}$ the analytical covariance matrix~\citep{Krause:2016jvl, DES:2020ypx}.

\textbf{\acs{BAO} and \acs{SNe Ia}} -- For \ac{DES} Y6 \ac{BAO}, we use the likelihood from~\citet{DES:2024pwq} or~\citet{DES:2026aht}, depending on whether the region overlapping with \ac{DESI} is included. It is provided as a $\chi^2$ table in the scaling parameter $\alpha \equiv (D_M(\Theta)/D_{M,\mathrm{fid}})(r_{d,\mathrm{fid}}/r_d(\Theta))$, i.e., the ratio of the comoving angular diameter distance $D_M$ to the sound horizon at the drag epoch $r_d$, relative to its value in a fiducial \textit{Planck} cosmology. For \ac{DESI} \ac{DR2} \ac{BAO}, we use the public likelihood~\citep{DESI:2025zgx} parameterized by the isotropic dilation $\alpha_{\mathrm{iso}} = (\alpha_\parallel \alpha_\perp^2)^{1/3}$ and the Alcock--Paczy\'nski parameter $\alpha_{\mathrm{AP}} = \alpha_\parallel/\alpha_\perp$, where $\alpha_{\parallel,\perp}$ are the line-of-sight and transverse scaling factors. For \ac{SNe Ia}, we assume the Gaussian likelihood from~\citet{DES:2025sig},
\begin{equation}
    \mathcal{L}\!\left(\mu_{\mathrm{fid}}\mid\Theta\right) \propto 
    \exp\!\left\{-\frac{1}{2}\,\Delta^i_\mu\,
    \mathrm{Cov}^{-1}_{ij}\,\Delta^j_\mu\right\},
\end{equation}
where $\Delta^i_\mu = \mu^i_{\mathrm{fid}} - \mu(z^i,\Theta)$ is the residual between observed and predicted distance moduli, and $\mathrm{Cov}_{ij}$ is the covariance matrix including statistical and systematic uncertainties~\citep{DES:2024hip}.

\textbf{Standard sirens} -- We adopt the spectral siren framework used by \ac{LVK}~\citep{LIGOScientific:2026uyd, LIGOScientific:2026ctl}. Detectors measure redshifted (detector-frame) masses; recovering source-frame masses requires the redshift, and hence a cosmology-dependent distance--redshift relation, through the factor $(1+z)$. Distinct features in the intrinsic mass spectrum (peaks, cutoffs, rate changes) act as anchors, linking the observed mass distribution to redshift and enabling cosmological inference without identifying individual hosts. Redshift information is recovered via Bayesian hierarchical inference~\citep{Mandel:2018mve, Vitale:2020aaz}, assuming \ac{CBC} parameters are drawn from a population distribution whose hyperparameters $\Phi$ are inferred jointly with the cosmological parameters $\Theta$ and marginalized over as nuisance parameters. For a catalog of $N_{\mathrm{det}}$ detections, the hierarchical likelihood reads
\begin{equation}
    \mathcal{L}\!\left(D_{\mathrm{GW}} \mid \Theta, \Phi\right) \propto 
    \prod_{i}^{N_{\mathrm{det}}} 
    \frac{\int \mathrm{d}\theta\; 
    \mathcal{L}(D_{\mathrm{GW},i}\mid\theta)\,
    \pi(\theta\mid \Theta, \Phi)}{\xi(\Theta, \Phi)},
    \label{eqn:siren}
\end{equation}
where $D_{\mathrm{GW},i}$ is the data for the $i$-th event, $\theta$ its single-event parameters (masses, spins, distance, and inclination), $\mathcal{L}(D_{\mathrm{GW},i}\mid\theta)$ the single-event likelihood, and $\pi(\theta\mid\Theta,\Phi)$ the population prior encoding the mass, spin, and redshift distributions. The term $\xi(\Theta,\Phi) = \int \mathrm{d}\theta\; P_{\mathrm{det}}(\theta)\,\pi(\theta\mid\Theta,\Phi)$ corrects for selection effects, with $P_{\mathrm{det}}(\theta)$ the probability that an event with parameters $\theta$ is detected~\citep{Mandel:2018mve}.
For GW170817, we obtain multi-messenger posteriors by importance-sampling the \ac{LVK} \ac{PE} samples~\citep{LIGOScientific:2018mvr} with an external prior on the inclination angle $\theta_{\mathrm{JN}}$ derived from superluminal jet observations~\citep{Mooley:2018qfh}, following the procedure of~\citet{Hotokezaka:2018dfi}.

\textbf{\acs{CMB}} -- The likelihood for the TT, TE, and EE power spectra takes the form
\begin{equation}
    \mathcal{L}\!\left(C_\ell \mid \Theta\right) \propto 
    \exp\!\left\{-\frac{1}{2}\sum_{\ell} 
    \Delta_\ell^\top\,\mathrm{Cov}^{-1}_\ell\,\Delta_\ell\right\},
\end{equation}
where $\Delta_\ell = C^{\mathrm{obs}}_\ell - C_\ell(\Theta)$, $C^{\mathrm{obs}}_\ell = (C_\ell^{TT}, C_\ell^{TE}, C_\ell^{EE})$ are the observed spectra, $C_\ell(\Theta)$ the theoretical predictions, and $\mathrm{Cov}_\ell$ the covariance matrix at each multipole.

\textbf{Cosmological inference and tension metrics} -- We perform parameter inference within a Bayesian framework using \ac{MCMC} sampling for each probe individually. Prior distributions \lvkir{(see Table~\ref{tab:priors})} are homogenized across all analyses to enable consistent posterior combination. We adopt flat priors on cosmological parameters following \ac{DES} choices~\citetext{\citetalias{DES:2026zjp}; \citealt{desy6-3x2pt}}, with widened priors on $\Omega_m$ and $H_0$ to accommodate the ranges explored in~\citet{LIGOScientific:2026uyd}. We further apply a \ac{BBN} consistency check on $\Omega_b h^2$, rejecting samples incompatible with the predicted light-element abundances. Nuisance parameters specific to each probe retain their fiducial priors from the respective data releases. Posterior distributions are sampled using the \textsc{Nautilus} nested sampling algorithm~\citep{Lange:2023ydq} as implemented in \textsc{CosmoSis}~\citep{Zuntz:2014csq}. We report constraints as \desirjm{median} $\pm$ 68\% credible intervals, and quantify constraining power via the figure of merit $\mathrm{FoM}_{\theta_1,\theta_2} = (\det\,\mathrm{Cov}(\theta_1,\theta_2))^{-1/2}$.

Probe combinations are constructed using the normalizing flows approach implemented in \textsc{tensiometer}~\citep{DES:2024jgw, Raveri:2024dph}\footnote{\url{https://github.com/mraveri/tensiometer}}, which reconstructs the joint posterior from individual probe posteriors assuming consistent priors and the absence of significant tension among inputs. We assess the consistency of posterior distributions over parameters $p$ obtained from two experiments, $A$ and $B$ \desirjm{when assuming $\Lambda$CDM model}. Following~\citet{Raveri:2018wln, Raveri:2021wfz}, we compute the probability density of the parameter difference,
\begin{equation}
    P(\Delta p) = \int P(p|A)\,
    P(p - \Delta p|B)\,\mathrm{d}p\,.
    \label{eqn:tension}
\end{equation}
In higher-dimensional parameter spaces, we construct density estimators of $P(p|A)$ and $P(p|B)$ using normalizing flows and evaluate the integral via Monte Carlo sampling, as implemented in \textsc{tensiometer}~\citep{Raveri:2021wfz}. In one dimension, this reduces to the parameter-difference estimator $\Delta p = |\langle p \rangle_A - \langle p \rangle_B| / \sigma_{AB}$, where $\sigma_{AB} = (\sigma_A^2 + \sigma_B^2)^{1/2}$. All tensions are reported in units of equivalent Gaussian standard deviations ($\sigma$). Datasets are considered consistent, and their posteriors combined, when the measured tension falls below $3\sigma$.

\section{Results}
\label{sec:results}

\begin{figure}[h]
    \centering
    \includegraphics[width=0.82\columnwidth]{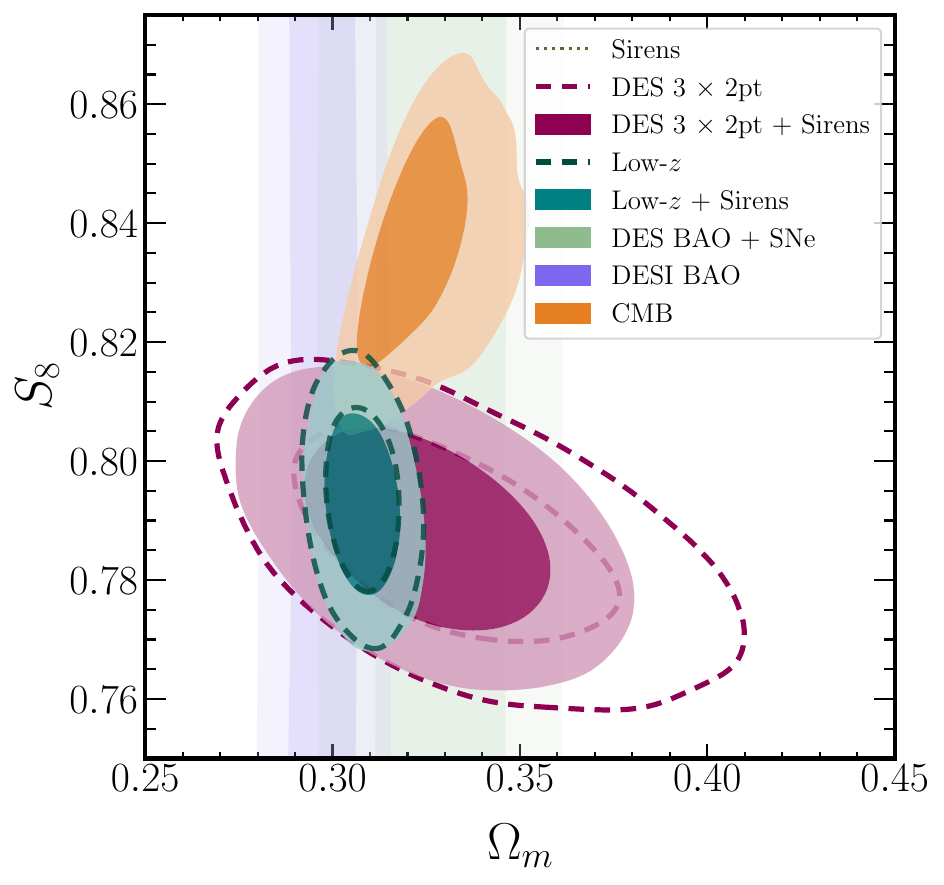}\\[2pt]
    \includegraphics[width=0.8\columnwidth]{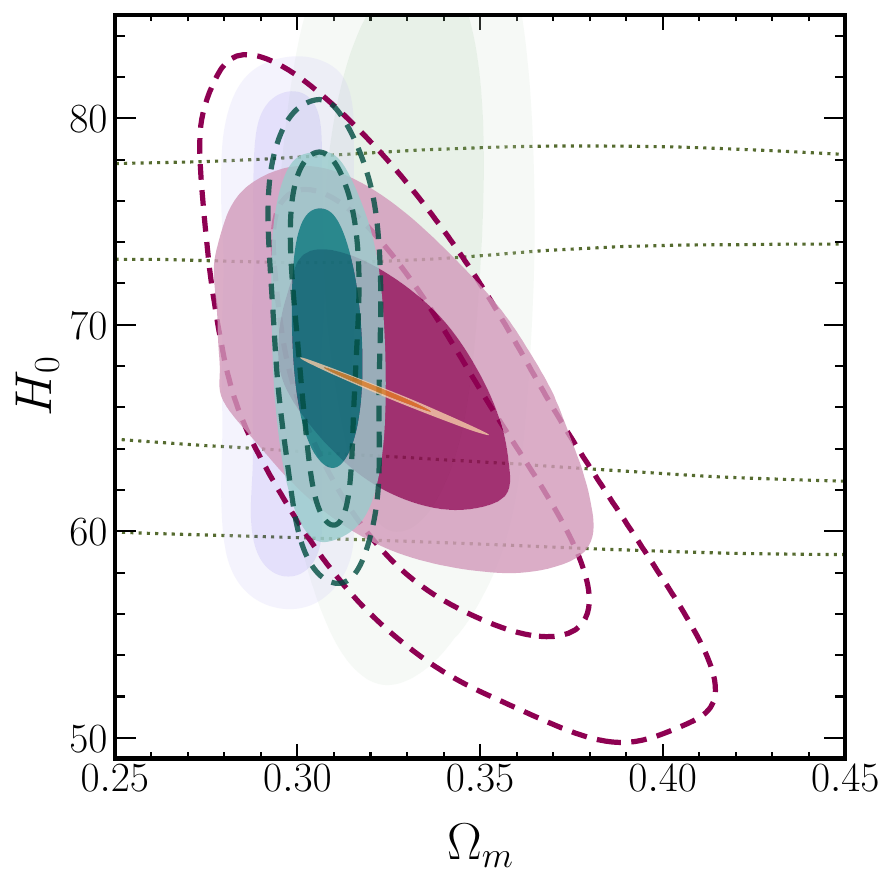}
    \caption{Marginalised posteriors in the $\Omega_m$--$S_8$ (top) and $H_0$--$\Omega_m$ (bottom) planes, showing 68\% and 95\% credible regions. Individual probes: DES Y6 3$\times$2pt (purple dashed), DES BAO + \acs{SNe Ia} (green), \acs{DESI} \acs{DR2} \acs{BAO} (violet), and \acs{GWTC-5.0} sirens (grey dotted, bottom panel; their 68\% and 95\% contours appear as near-vertical lines). Combinations: 3$\times$2pt + sirens (purple), low-$z$ (teal dashed), low-$z$ + sirens (teal), and \acs{CMB} (\textit{Planck} + \acs{ACT} + \acs{SPT}-3G; orange). Low-$z$ = 3$\times$2pt + DES BAO + DES \acs{SNe Ia} + \acs{DESI} BAO.}
    \label{fig:posteriors}
\end{figure}

Figure~\ref{fig:posteriors} (top) shows constraints in the $S_8$--$\Omega_m$ plane. \ac{DES} Y6 3$\times$2pt alone measures $S_8$ at the 1.5\% level; when combined with \ac{LVK} standard sirens, the figure of merit in this plane improves by \ver{$\sim$30\%}. Although sirens carry no direct sensitivity to structure growth, the gain arises from the complementary degeneracy directions of the two probes \ver{in the $\Omega_m - H_0$ plane}. When the full low-$z$ combination is considered, adding standard sirens leaves both the constraining power and the consistency with \ac{CMB} unchanged, as the geometric information in sirens is already captured by \ac{BAO} and \ac{SNe Ia}. The 3$\times$2pt + standard sirens combination is subject to systematics largely uncorrelated with those of more established probes such as \ac{BAO} or galaxy clusters, providing an independent cross-check. The bottom panel of Figure~\ref{fig:posteriors} shows marginalized posteriors in the $H_0$--$\Omega_m$ plane.  Standard sirens play a key role here: combined with 3$\times$2pt, they constrain the Hubble constant to \ver{6.2\%} (see Table~\ref{tab:constraints}), \ver{improving minimally} on the measurement of \cite{Andrade-Oliveira:2026jjm}.

Additionally, we report cosmology constraints from the most precise combination of low-$z$ probes including standard sirens:
\ver{
\begin{equation}
    S_8 = 0.793 \pm 0.010, \; \Omega_m = 0.308^{+0.007}_{-0.006}, \; H_0 = 69.4^{+4.1}_{-4.4} \,\mathrm{km}\,\mathrm{s}^{-1}\,\mathrm{Mpc}^{-1},
\end{equation}
}
reaching a similar precision on the clustering amplitude and the matter abundance as the low-$z$ combination presented in \citet{desy6-3x2pt} (see Table~\ref{tab:constraints} for all combinations).

\begin{figure}
    \centering
    \includegraphics[width=0.9\columnwidth]{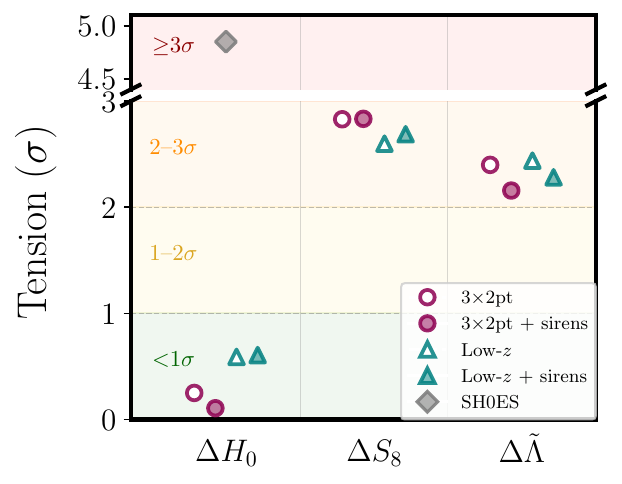}
    \caption{Consistency of dataset combinations with \acs{CMB} in $H_0$, $S_8$ (1D), and the shared parameter space $\tilde{\Lambda}$. Open (filled) markers exclude (include) \acs{GWTC-5.0} sirens; circles are 3$\times$2pt and triangles low-$z$. Shading marks the $<1\sigma$, $1$--$2\sigma$, $2$--$3\sigma$, and ${>}3\sigma$ regions.}
    \label{fig:tension_forest}
\end{figure}

Finally, Figure~\ref{fig:tension_forest} and Table~\ref{tab:tensions} show tension metrics with respect to \ac{CMB}. In 1D, \ac{DES} Y6 3$\times$2pt finds $S_8$ in $2.83\sigma$ tension with \ac{CMB}, consistent with prior \ac{DES} analyses; this tension decreases mildly to \ver{$2.6\sigma$} for the full low-$z$ combination, and is stable to the inclusion of standard sirens at the \ver{$0.09\sigma$} level. In the $\Omega_m$--$S_8$ plane, the tension with \ac{CMB} ranges \ver{from $2.15\sigma$ to $2.60\sigma$} across the combinations considered, with none exceeding $3\sigma$. At current \ac{GWTC-5.0} sensitivity, standard sirens do not resolve whether the data prefer early- or late-Universe values of $H_0$; nevertheless, the precision gain shows this probe is already a meaningful contributor, with $\Delta H_0$ shifting by at most \ver{${\sim}0.1\sigma$} when sirens are included --- a role of precision improvement rather than tension resolution, set to grow as the \ac{LVK} catalog expands. Moving to the full parameter space --- the set of parameters shared across all analyses, $\tilde{\Lambda} \equiv \{\Omega_m, S_8, \Omega_b, h, n_s, \sum m_\nu\}$ --- the consistency of both 3$\times$2pt and the full low-$z$ combination with \ac{CMB} remains stable, ranging \ver{from $2.16\sigma$ to $2.44\sigma$} regardless of whether standard sirens are included.

\section{Conclusions}
\label{sec:conclusions}

We presented the first joint multiprobe $\Lambda$CDM analysis combining low-$z$ \ac{EM} datasets with \ac{LVK}~\ac{GWTC-5.0} standard sirens.

Three main results emerge from this analysis. First, standard sirens improve the figure of merit in the $\Omega_m$--$S_8$ plane by $\sim30\%$ when added to 3$\times$2pt, owing to their complementary parameter degeneracies in $\Omega_m-H_0$. Second, the combined 3$\times$2pt + standard sirens dataset constrains the Hubble constant at the 6.2\% precision level, \ver{slightly} improving on the measurement in~\cite{Andrade-Oliveira:2026jjm} thanks to the addition of \ac{GWTC-5.0} data, and highlighting the increasing weight of spectral sirens relative to the single bright siren in constraining $H_0$. Third, we deliver the first $\Lambda$CDM tension assessment including \acp{GW}; current \ac{GWTC-5.0} standard sirens lack the constraining power to resolve either the $H_0$ tension or the mild $S_8$ discrepancy, but this analysis demonstrates that standard sirens have reached the same level of maturity as other geometric probes and can be routinely incorporated into multiprobe analyses. Given the projected \ac{LVK} observing campaigns and synergies with photometric surveys, substantial gains in constraining power are anticipated.

We demonstrate the feasibility of such analyses and recommend that upcoming surveys --- \ac{LSST}, \textit{Euclid}, and \textit{Roman} --- incorporate standard sirens in their multiprobe pipelines.

\begin{acknowledgements} 
Funding for this work is provided by the University of Zurich (UZH) and the Swiss National Science Foundation (SNSF) under grant number 10002981. DL is supported by the UZH Postdoc Grant, grant no. [K-72341-01-01]. This material is based upon work supported by NSF’s LIGO Laboratory which is a major facility fully funded by the National Science Foundation. This project used public archival data from the Dark Energy Survey (DES). Funding for the DES Projects has been provided by the DOE and NSF(USA), MEC/MICINN/MINECO(Spain), STFC(UK), HEFCE(UK). NCSA(UIUC), KICP(U. Chicago), CCAPP(Ohio State), MIFPA(Texas A\&M), CNPQ, FAPERJ, FINEP (Brazil), DFG(Germany) and the DES Collaborating Institutions. Based in part on observations at Cerro Tololo Inter-American Observatory, NOIRLAB, which is operated by the Association of Universities for Research in Astronomy (AURA) under a cooperative agreement with the NSF. The authors thank Surhud More, Juan Mena-Fernandez, and Hugo Camacho for their thorough internal review, which improved the robustness of this work.
\end{acknowledgements}

\tiny{\textit{Author Contributions.} DSC led the analysis and drafted the manuscript. FAO contributed to the analysis with spectral sirens posteriors. FAO, DL, MSS contributed to the interpretation of results and reviewed the manuscript. The remaining authors contributed to DECam construction and data collection, processing, and calibration; developed methods, codes, and simulations; ran pipelines and validation tests; and supported the science analysis.}

\bibliographystyle{aa}
\bibliography{refs}

\printaffiliations 

\appendix

\begin{table*}[h]
\section{Prior distribution}
    \centering
    \caption{Priors on cosmology parameters.}
    \label{tab:priors}
    \begin{tabular}{lr}
        \hline \hline
        Parameter & Prior \\
        \hline
        $\Omega_{\rm m}$ & [0.0, 1.0] \\
        $A_{\rm s} \times 10^9$ & [0.5, 5] \\
        $h$ & [0.1, 2.0] \\
        $\Omega_{\rm b}$ & [0.03, 0.07] \\
        $n_{\rm s}$ & [0.93, 1.00] \\
        $\Sigma m_\nu$ [eV] & [0.06, 0.6] \\
        \hline \hline
    \end{tabular}
    \tablefoot{All priors are flat, denoted by [min, max]. Astrophysical and calibration parameters are sampled with the priors defined in the original analysis of each dataset.}
\end{table*}

\begin{table*}[h]
\section{Tension metrics}
    \small
    \setlength{\tabcolsep}{3pt}
    \caption{Consistency metrics with \ac{CMB}, in units of $\sigma$.}
    \label{tab:tensions}
    \centering
    \begin{tabular}{lcccc}
    \hline\hline
    \textbf{Dataset} & $\Delta H_0$ & $\Delta S_8$ & $\Delta(\Omega_m,S_8)$ & $\Delta\tilde{\Lambda}$\\
    \hline
    3$\times$2pt            & 0.25 & 2.83 & 2.48 & 2.40 \\
    3$\times$2pt + Sirens   & 0.11 & 2.83 & 2.60 & 2.16 \\
    Low-$z$                 & 0.59 & 2.60 & 2.15 & 2.44 \\
    Low-$z$ + sirens        & 0.61 & 2.69 & 2.26 & 2.28 \\
    SH0ES                   & 4.85 & --   & --   & --   \\
    \hline
    \end{tabular}
    \tablefoot{We evaluate consistency in $S_8$, $H_0$, and the full shared parameter space $\tilde{\Lambda} \equiv \{\Omega_m, S_8, \Omega_b, h, n_s, \sum m_\nu\}$. One-dimensional tensions are computed as parameter differences.}
\end{table*}

\begin{table*}[b]
\section{Cosmology constraints}
    \setlength{\extrarowheight}{7pt}
    \caption{Marginalized $\Lambda$CDM parameter constraints.}
    \label{tab:constraints}
    \resizebox{\textwidth}{!}{%
    \begin{tabular}{lccccccc|cc}
    \hline\hline
    \textbf{Dataset} & $S_8$ & $\Omega_{\mathrm{m}}$ & $\sigma_8$ & $h$ & $\Omega_{\mathrm{b}} \times 10^2$ & $n_{\mathrm{s}}$ & $\sum m_{\nu}$ (eV) & $\mathrm{FoM}_{S_8,\Omega_m}$ & $\mathrm{FoM}_{H_0,\Omega_m}$ \\ [0.4em]
    \hline
    $3\times2$pt & $0.787^{+0.012}_{-0.012}$ & $0.331^{+0.031}_{-0.026}$ & $0.749^{+0.039}_{-0.040}$ & $0.644^{+0.078}_{-0.066}$ & $4.45^{+0.97}_{-0.86}$ & $0.965^{+0.023}_{-0.023}$ & $-$ & 3436 & 7 \\ [0.4em]
    $3\times2$pt + Sirens & $0.788^{+0.011}_{-0.011}$ & $0.325^{+0.022}_{-0.021}$ & $0.758^{+0.031}_{-0.030}$ & $0.670^{+0.043}_{-0.040}$ & $4.68^{+0.84}_{-0.81}$ & $0.962^{+0.024}_{-0.022}$ & $-$ & 4699 & 14 \\ [0.4em]
    Low-$z$ & $0.794^{+0.010}_{-0.011}$ & $0.308^{+0.007}_{-0.006}$ & $0.784^{+0.014}_{-0.014}$ & $0.707^{+0.051}_{-0.071}$ & $5.08^{+0.51}_{-1.00}$ & $0.967^{+0.020}_{-0.023}$ & $-$ & 15274 & 28 \\ [0.4em]
    Low-$z$ + Sirens & $0.793^{+0.010}_{-0.010}$ & $0.308^{+0.007}_{-0.006}$ & $0.783^{+0.014}_{-0.014}$ & $0.694^{+0.041}_{-0.044}$ & $4.94^{+0.45}_{-0.62}$ & $0.968^{+0.021}_{-0.023}$ & $-$ & 15181 & 37 \\ [0.4em]
    CMB & $0.836^{+0.013}_{-0.013}$ & $0.323^{+0.011}_{-0.009}$ & $0.807^{+0.01}_{-0.014}$ & $0.668^{+0.006}_{-0.008}$ & $5.03^{+0.12}_{-0.09}$ & $0.97^{+0.004}_{-0.004}$ & $<0.24$ & 8380 & 1190 \\ [0.4em]
    \hline\hline
    \end{tabular}%
    }
    \tablefoot{Values reported as median and 68\% credible intervals for $S_8$, $\Omega_{\mathrm{m}}$, $\sigma_8$, $h$, $\Omega_{\mathrm{b}}$, and $n_{\mathrm{s}}$; 95\% upper limit on $\sum m_\nu$, where a dash denotes a prior-dominated posterior for which no meaningful limit is reported; and figures of merit in the $S_8$--$\Omega_{\mathrm{m}}$ and $H_0$--$\Omega_{\mathrm{m}}$ planes.}
\end{table*}

\begin{figure*}[t]
\section{Hubble constant posterior distribution}
    \centering
    \includegraphics[width=0.65\linewidth]{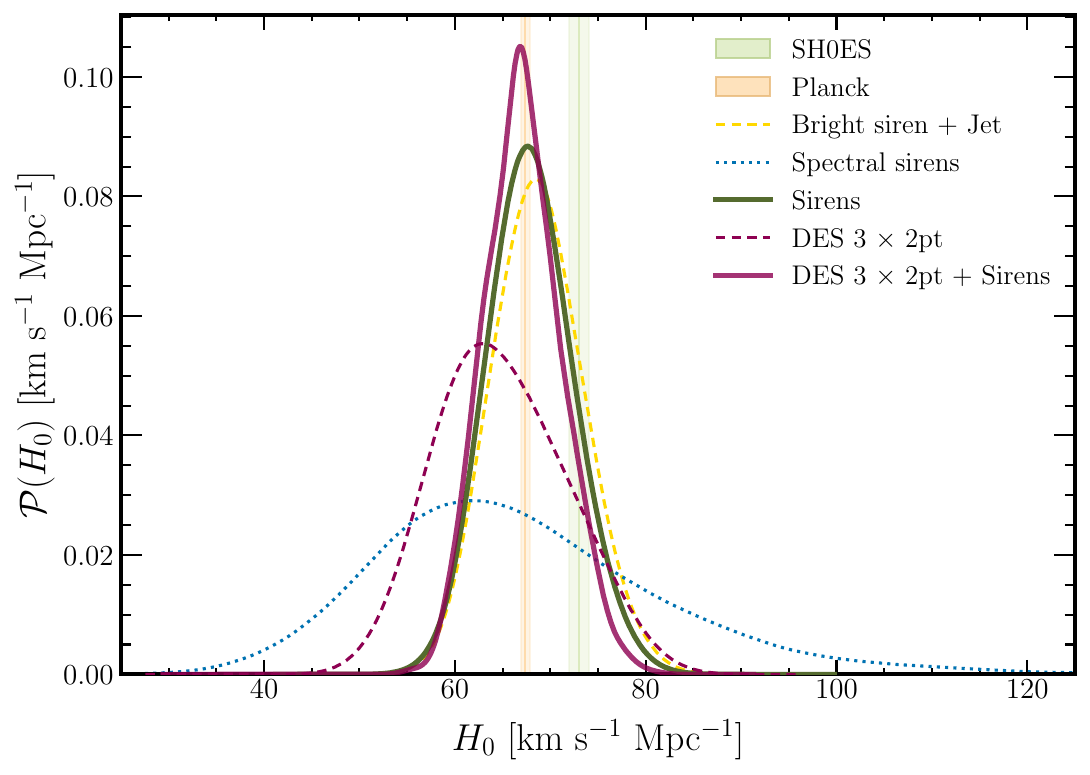}
    \caption{\label{fig:h0}Marginalized $H_0$ constraints from GW170817 + jet (Bright siren + Jet, dashed yellow), spectral sirens (dotted blue), both data sets are combined into Sirens (dark green); \ac{DES} 3 $\times$ 2pt (dashed purple), and \ac{DES} 3 $\times$ 2pt + Sirens (purple). The orange and green bands are \ac{CMB} constraints from \textit{Planck}~\citep{Planck:2018vyg} and \ac{SNe Ia} from SH0ES~\citep{Riess:2021jrx} constraints at the 68\% credible interval.}
\end{figure*}

\end{document}